\documentclass[showpacs,preprintnumbers,amsmath,amssymb]{revtex4}
\usepackage{graphicx}% Include figure files
\usepackage{dcolumn}% Align table columns on decimal point
\usepackage{bm}% bold math

\newcommand{\hub}{{\cal H}}
\newcommand{\z}{{\cal Z}}

\newcommand{\s}{\sigma}
\newcommand{\p}{{\prime}}
\newcommand{\pp}{{\prime\prime}}
\newcommand{\ppp}{{\prime\prime\prime}}

\begin{document}

%\preprint{APS/123-QED}

\title{Evolution of Cosmological Perturbations in the Presence of Primordial
Magnetic Fields}% Force line breaks with \\

\author{Kazuhiko Kojima}
\email{kazuhiko.kojima@nao.ac.jp}
\altaffiliation[Also at ]{Division of Theoretical Astrophysics, National Astronomical
Observatory, Tokyo 181-8588, Japan
}%
\affiliation{%
Department of Astronomy, University of Tokyo, Tokyo 113-0033, Japan
}%

\author{Kiyotomo Ichiki}%
\affiliation{Department of Physics and Astrophysics, Nagoya
University, Nagoya 464-8602, Japan
}%

\date{\today}% It is always \today, today,
% but any date may be explicitly specified

\begin{abstract}
 Possible existence of the primordial magnetic fields has
 affected the structure formation of the universe. 
 In this paper it is shown that the initial conditions for density
 perturbations with magnetic fields derived in previous works are
 inconsistent with Einstein equations. 
 We find that this inconsistency arises due to the unwanted cancellation
 of contributions from the magnetic fields and primordial radiations.  
 A complete set of equations and consistent initial conditions in the
 long wavelength limit are given with an explicit derivation 
 in the covariant approach with CDM frame, by newly taking into account a
 non-relativistic matter contribution in the radiation dominated era.  
 By solving these equations numerically, we derive the angular spectrum
 of cosmic microwave background anisotropies and the matter power
 spectrum with magnetic fields.
 We find that the amplitude of
 the angular power spectrum of CMB anisotropies can alter at most a
 order of magnitude at $l \lesssim 4000$ 
 compared with the previous results in the literature. 
\end{abstract}
\maketitle

%--------------------------------------------------------------------------
\section{Introduction}
%--------------------------------------------------------------------------
There are mounting evidences that the large scale magnetic fields are
present in various objects in the universe. Not only galaxies,
but also clusters of galaxies contain their own magnetic
fields with the field strength of $\sim 10^{-6}$ Gauss and the
coherence length of $1 - 10$ kpc (for a review,
\cite{2002RvMP...74..775W}). Furthermore, there have been some 
observations which indicate that they exist even in larger scales, such
as in superclusters \cite{2006ApJ...637...19X}.  

%--------------------------------------------------------------------------
Yet the origin of such large scale magnetic fields is still a
matter of debate. Magnetic fields in spiral galaxies are assumed to be
continuously generated and maintained by dynamo mechanism
\cite{1971ApJ...163..255P}. However, one still needs to explain the
origin of seed fields necessary for dynamo action to take place.
Astrophysical origins of such seed fields, often involving stellar
activities or the 
Biermann battery in non-adiabatic processes
\cite{Biermann50,2000ApJ...539..505G,2005ApJ...633..941H}, may explain the 
strength and the total amount of magnetic fields with help from the dynamo
mechanism. Their coherence scales are, however, much smaller than those
of intergalactic magnetic fields and thus magnetic fields generated from
these mechanisms could not be directly the origin of large-scale
magnetic fields. 

%----------------------------------------------------------------------
On the other hand, primordial origins, often related with
inflation
\cite{Ratra:1991bn,Turner:1987bw,Bamba:2003av,2004PhRvD..70d3004P,2005PhRvD..71j3509A,2008JCAP...01..025M}
or second order effects through cosmological vector modes
\cite{1970MNRAS.147..279H,Hogan:2000gv,2005MNRAS.363..521G,2004APh....21...59B,2006Sci...311..827I,2005PhRvL..95l1301T,2008arXiv0805.0169M,2007PhRvD..75j3501K},
have no difficulty in accounting for the length of coherence. The
observational facts that there exist significant magnetic fields in
objects at large redshift may support the hypothesis of primordial origin of the
large scale magnetic fields \cite{1992ApJ...387..528K,2008Natur.454..302B}. If this is the case, it is expected that
primordial magnetic fields should have formed imprints in the
anisotropies of cosmic microwave background (CMB) through their stress
energy tensor and their Lorentz force on the baryon-photon
fluid before cosmological recombination. Therefore, it is important to 
develop the cosmological perturbation theory with primordial magnetic
fields in order to search for signs of magnetic fields in the observed
CMB maps.

%--------------------------------------------------------------------------
In recent years the effects of stochastic primordial magnetic fields on the
evolution of cosmological perturbations have been developed
independently by several authors (for a review
\cite{2007PhR...449..131B}). The scalar type perturbations, which 
is related to density fluctuations, have been considered by 
\cite{1996PhLB..388..253A,2000PhRvD..61h3519T,2000PhRvD..62h3509K,2006MNRAS.368..965T,2006PhRvD..74l3518Y,2008PhRvD..77f3003G,2008arXiv0811.0230P}.
The vector type perturbations, which would give the most dominant
contribution to CMB anisotropies at small angular scales, have been
studied by
\cite{1998PhRvL..81.3575S,2002PhRvD..65l3004M,2002MNRAS.335L..57S,2003MNRAS.344L..31S,2004PhRvD..70d3011L,2005ApJ...625L...1Y,2008PhRvD..78d5010K}, and
also the tensor type perturbations
\cite{2002PhRvD..65l3004M,2000PhRvD..61d3001D}. All of these studies
suggest that, from the currently available CMB data,  the amplitude of
primordial stochastic magnetic fields should be at most a few times
$10^{-9}$ Gauss or below at the relevant scales \cite{2006ApJ...646..719Y}. 

%--------------------------------------------------------------------------
In the present paper we reconsider the scalar type cosmological
perturbations with primordial magnetic fields. Within the standard
cosmological perturbation theory, an initial condition of the
perturbation Fourier mode with wavenumber $k$ is set when the mode is well
outside the horizon ($k\tau \ll 1$ with 
$\tau$ being conformal time) and when the universe is deep in the
radiation dominated era, neglecting non-relativistic matter
contributions (for example, see \cite{1995ApJ...455....7M}). Following
this standard practice the initial conditions of density perturbations
with primordial magnetic fields have been derived
\cite{2008PhRvD..78b3510F,2008PhRvD..77l3001G,2004PhRvD..70l3507G,2008PhRvD..77f3003G,2008PhRvD..78l3001Y}. 
We find, however, that this procedure does not give us a consistent
initial conditions in the presence of magnetic fields, because of the
unwanted cancellation of contributions from the magnetic fields and
primordial radiations. 
This cancellation makes the system unstable and
violates the constraint of perturbed Einstein equations to be satisfied.
As we shall show below, this difficulty can be removed by considering
the significant contributions from the non-relativistic matter at initial
conditions. 

%--------------------------------------------------------------------------
The paper is organized as follows.
In Section II, we set up basic equations for the perturbation theory
including primordial magnetic fields.
We adopt the covariant approach to derive equations while equations in
Refs. \cite{1995ApJ...455....7M,2008PhRvD..77f3003G,2008arXiv0811.0230P,2008PhRvD..78b3510F}
are derived in a conventional synchronous gauge.
The anisotropic stress and the Lorentz force of the 
magnetic field are defined and the equation of motion for baryons is derived.
We also define the spectrum of the magnetic field.
In Section III, we point out an inconsistency in the previous works and
derive the initial condition of the purely magnetic mode
including the non-relativistic matter contribution. 
In Section IV we show the numerical calculation of CMB and matter power
spectra.  Finally, we conclude this work.

%--------------------------------------------------------------------------
\section{equations}
\subsection{Basic equations}
Here we set up equations.  In what follows we take the covariant approach
with CDM frame to eliminate the gauge  
freedom \cite{1992ApJ...395...54D,1999ApJ...513....1C,2002PhRvD..65j4012L}.
In this frame, we define variables on the supersurface orthogonal to the
CDM 4-velocity $u_\mu$. 
Then one can define the anisotropic expansion rate (shear) $\sigma$ and
the inhomogeneous expansion rate $\z$ from the covariant derivative of
$u_\mu$. In the scalar mode, we can neglect vorticity.
In addition, we introduce the Weyl tensor, which is the traceless part 
of the Riemann curvature tensor.
The Weyl tensor vanishes in the background FRW spacetime.
Since the magnetic part of the Weyl tensor is negligible in the scalar mode,
we define the electric part of the Weyl tensor as $\Phi$.
Linearizing the Bianchi identities and Ricci identities, we obtain the following
equations  
for $\Phi$, $\s$ and $\z$.
Three propagation equations for $\Phi$, $\s$ and $\z$ are:
\begin{eqnarray}
&&\Phi^\p+\hub\Phi+\frac{1}{2k}\kappa\rho a^2\left(\tilde{\gamma}\sigma+R_fq_f\right)+\frac{\hub}{2k^2}\kappa\rho a^2(3\tilde{\gamma}-1)R_f\pi_f-\frac{1}{2k^2}\kappa\rho a^2R_f\pi_f^\p=0~,\label{chap3:1}\\
&&\sigma^\p+\hub\sigma+k\Phi+\frac{1}{2k}\kappa \rho a^2R_f\pi_f=0~,\label{chap3:2}\\
&&\z^\p+\hub\z+\frac{1}{2k}\kappa\rho a^2 R_f(\Delta_f+3\delta P_f)=0~.\label{chap3:5}
\end{eqnarray}
Two constraint equations are:
\begin{eqnarray}
&&\frac{2}{3}(\z-\s)+\frac{1}{k^2}\kappa\rho a^2R_fq_f=0~,\label{chap3:3}\\
&&2\Phi-\frac{1}{k^2}\kappa\rho a^2(R_f\Delta_f+R_f\pi_f)-\frac{3\hub}{k^3}\kappa\rho a^2R_fq_f=0~.\label{chap3:4}
\end{eqnarray}
The prime "$~^\p~$" denotes the derivative with respect to 
the conformal time $\tau$ and $\hub\equiv a^\p/a$, where $a$ is the
scale factor. Here we defined $\tilde{\gamma}$ as $p=(\tilde{\gamma}-1)\rho$, where
$p$ is the total pressure and $\rho$ is total energy density, and $\kappa\equiv 8\pi G$.
The subscript "$~_f~$" means the sum of photon ($\gamma$),
neutrino ($\nu$), baryon ($b$), CDM ($c$) and magnetic field ($B$). 
The density fluctuation $\Delta_f$, heat flux $q_f$ and
anisotropic stress $\pi_f$ are normalized with their energy density 
$\rho_f$ except for the magnetic field. For the magnetic field variables
we normalize $\Delta_B$ and $\pi_B$ with the photon energy density
$\rho_\gamma$ because 
we consider the case that the magnetic field does not contribute to the
background spacetime.
Then the total energy density $\rho$ is
$\rho=\rho_\gamma+\rho_\nu+\rho_b+\rho_c$. 
Here we define the energy fraction $R_f$ as $R_\gamma=\rho_\gamma/\rho$,
$R_\nu=\rho_\nu/\rho$, $R_b\tau=\rho_b/\rho$, $R_c\tau=\rho_c/\rho$ and
$R_m\equiv \frac{3}{4}(R_b+R_c)$.  
In the deep radiation dominated era, the background energy densities of 
baryon and CDM are negligible. In this epoch, 
$\rho\simeq\rho_\gamma+\rho_\nu$,  $R_\gamma+R_\nu\simeq1$ and all $R_f$'s 
are constant. Note that $R_b$ and $R_c$ have dimension of (length)$^{-1}$.
Note that in the metric perturbation approach, there are four, not five,
equations as shown in Ref. \cite{1995ApJ...455....7M}. This is because
our variables $\Phi$, $\s$ and $\z$ are not independent from each other. 

Next we introduce the stochastic magnetic field and evolution equations
for each component.  In this work, we assume that the magnetic field can
be treated as the first order quantity and does not contribute to the
background evolution.   
Then the perturbed Einstein equation is described as 
\begin{eqnarray}
\delta G^\mu_{~\nu}=8\pi G(\delta T^\mu_{~\nu}+T^{~~\mu}_{B~\nu})~.
\end{eqnarray} 
Here $T^{~\mu}_{B~\nu}$ is the energy-momentum tensor for the magnetic field:
\begin{eqnarray}
&&T^{~~0}_{B~0}(x)=-\frac{B^2(x)}{8\pi a^4}~,\\
&&T^{~~i}_{B~j}(x)=\frac{1}{4\pi a^4}\left(\frac{B^2(x)}{2}\delta^i_{~j}-B^i(x)B_j(x)\right)~,
\end{eqnarray}
where $B_i(x)$ is the magnetic field strength at present time and we
assume that the conductance of the universe is infinite, 
i.e. $E_i=0$.
We decompose the space-space part of energy-momentum tensor as 
\begin{eqnarray}
T^{~~i}_{B~j}(x)=\frac{1}{3}\frac{B^2(x)}{8\pi a^4}\delta^i_{~j}+\frac{1}{4\pi a^4}\left(-B^i(x)B_j(x)+\frac{1}{3}\delta^i_{~j}B^2(x)\right)~,
\end{eqnarray}
where the first term in the r.h.s. is the trace part and second term is
the traceless part. The traceless part is the anisotropic stress. 
In the Fourier space, the energy density and anisotropic
stress of the magnetic field are defined as 
\begin{eqnarray}
&&\rho_\gamma\Delta_B=-T^{~~0}_{B~0}(k)=\delta_i^{~j}T^{~~i}_{B~j}(k)~,\\
&&\rho_\gamma\pi_B=-\frac{3}{2}\left(\hat{k}_i\hat{k}^j-\frac{1}{3}\delta_i^{~j}\right)T^{~~i}_{B~j}(k)~.
\end{eqnarray}
Since $T^{~~\mu}_{B~\nu}(k)$ and $\rho_\gamma$ are proportional to $a^{-4}$, $\Delta_B$ and $\pi_B$ are
constant. 
Then $T^{~~i}_{B~j}(k)$ is expressed as
\begin{eqnarray}
T^{~~i}_{B~j}(k)=\frac{1}{3}\delta^{i}_{~j}\rho_\gamma\Delta_B-\left(\hat{k}^i\hat{k}_j-\frac{1}{3}\delta^i_{~j}\right)\rho_\gamma\pi_B~.
\end{eqnarray}
If the magnetic field exists, the time evolution of the baryon fluid is
affected by the Lorentz force. The energy conservation for baryon is
described as
\begin{eqnarray}
&&\delta T^{~\mu}_{b~\nu;\mu}+T^{~\mu}_{B~\nu;\mu}=0~,\\
&&\delta T^{~\mu}_{b~i;\mu}=\rho_b( q_{bi}^\p+\hub q_{bi})+\frac{4}{3}an_e\sigma_T\rho_\gamma\left(q_{bi}-\frac{3}{4} q_{\gamma i}\right)~,\\
&&T^{~~\mu}_{B~i;\mu}=\frac{1}{4\pi a^4}(\nabla\times {\bf B})\times{\bf
 B}=\frac{1}{4\pi a^4}
\left(({\bf B}\cdot\nabla){\bf B}-\frac{1}{2}\nabla B^2\right)=-\frac{1}{3}\frac{\nabla_j B^2}{8\pi a^4}\delta^j_{~i}-\frac{\nabla_j}{4\pi a^4}\left(-B^jB_i+\frac{1}{3}\delta^j_{~i} B^2\right)~,
\end{eqnarray}
where $n_e$ is the number density of free electrons and $\sigma_T$ is
the Thomson cross section. Here we have neglected  
the baryon pressure. Finally, we obtain the equation of motion for
baryon in the Fourier space: 
\begin{eqnarray}
&&q_b^\p+\hub q_b+an_e\sigma_TR\left(q_b-\frac{3}{4}q_\gamma\right)=-\frac{3}{4}kRL~,\nonumber\\
&&L\equiv\frac{1}{3}(-\Delta_B+2\pi_B)~,\label{chap3:b1}
\end{eqnarray}
where $R\equiv 4\rho_\gamma/(3\rho_b)$.
The Lorentz force term does not vanish after the recombination
because residual free electrons and ions  
still interact with neutral atoms, and thus these particles move
together. As we see later, the Lorentz force have a large effect on the 
growth of curvature perturbation after the recombination.
Under the circumstances where the electric field is negligible, the
continuity equation for baryon is written in the same manner as the
standard one: 
\begin{eqnarray}
&&\Delta_b^\p+k(\z+q_b)=0~.\label{chap3:B}
\end{eqnarray}

Photons are coupled with baryons through Thomson scattering.
The zero and first moments of Boltzmann equation for photons are
\begin{eqnarray}
&&\Delta_\gamma^\p=-k\left(\frac{4}{3}\z+q_\gamma\right)~,\label{chap3:bg0}\\ 
&&q_\gamma^\p=\frac{k}{3}(\Delta_\gamma-2\pi_\gamma)+an_e\sigma_T\left(\frac{4}{3}q_b-q_\gamma\right)~. \label{chap3:bg1}
\end{eqnarray}
In the early epoch at which one should impose the initial conditions,
the anisotropic stress and more higher multipoles of photons are
negligible. 

Before the recombination, baryons and photons are tightly
coupled, so that $q_b\simeq 3q_\gamma/4\equiv v_{\gamma b}$. 
In the tight-coupling epoch, Eqs. (\ref{chap3:bg1}) and (\ref{chap3:b1}) are combined to
\begin{eqnarray}
&&v_{\gamma b}^\p+\frac{\hub}{1+R}v_{\gamma b}=\frac{k}{4}\frac{R}{1+R}(\Delta_\gamma-3L)~.\label{chap3:photb}
\end{eqnarray}

Other particle species such as neutrino and CDM are treated as
collisionless particles.  Since neutrinos are relativistic, their
evolution should be followed by solving the collisionless Boltzmann
equations:  
\begin{eqnarray}
&&\Delta_\nu^\p=-k\left(\frac{4}{3}\z+q_\nu\right)\label{chap3:bn0}~,\\
&&q_\nu^\p=\frac{k}{3}(\Delta_\nu-2\pi_\nu)~, \label{chap3:bn1}\\
&&\pi_\nu^\p=k\left(\frac{2}{5}q_\nu-\frac{3}{5}G^{(3)}_\nu\right)+\frac{8}{15}k\sigma\label{chap3:bn2}~,\\
&&G^{(3)\p}_\nu=\frac{3}{7}k\pi_\nu\label{chap3:bn3}~,
\end{eqnarray}
where we set $G_\nu^{(l)}=0~(l>3)$.
CDM can be treated as a nonrelativistic perfect fluid and does not have
velocity in our frame, $q_c=0$. Its energy perturbation evolves as 
\begin{eqnarray}
\Delta_c^\p+k\z=0~.\label{chap3:c0}
\end{eqnarray}

\subsection{Spectrum of the magnetic field}
To calculate CMB anisotropies generated from stochastic primordial
magnetic fields, we need to specify the spectrum of them.
It is shown in Refs. \cite{2002PhRvD..65l3004M,2005PhRvD..71j3006K}
that the magnetic field is damped in small scales $k>k_D$, where $k_D$
is the wavenumber of damping scale. 
Here we assume that the magnetic field has power-law spectrum in $k<k_D$
in the same manner as in previous works
\cite{2002PhRvD..65l3004M,2004PhRvD..70d3011L,2007PhRvD..75b3002K},  
\begin{eqnarray}
\langle B_i({\bf k})B_j^\ast({\bf k}^{\prime})\rangle&=&(2\pi)^3\frac{P_{ij}}{{2} }Ak^{n_B}\delta({\bf k}-{\bf k}^{\prime})~,
\end{eqnarray}
with $P_{ij}\equiv\delta_{ij}-\hat{k}_i\hat{k}_j$, which is the divergence free condition of the magnetic field.
There are some ways to define the amplitude of the magnetic fields as
shown in Ref.  
\cite{2008PhRvD..78b3510F}. In this work,
we define the amplitude of the magnetic field $B_\lambda$ by smoothing
at $\lambda=1{\rm Mpc}$ with Gaussian window function in Fourier space, 
\begin{eqnarray}
B_\lambda^2\equiv\frac{1}{(2\pi)^3}\int d^3kAk^{n_B}\exp{(-\lambda^2k^2)}~.
\label{chap3:b-l}
\end{eqnarray}
Integrating Eq. (\ref{chap3:b-l}),
one obtains the two-point correlation of magnetic field,
\begin{eqnarray}
\langle B_i({\bf k})B_j^\ast({\bf k}^{\prime})\rangle&=&(2\pi)^3P_{ij}\frac{(2\pi)^{n_B+5}B_\lambda^2}{2\Gamma(\frac{n_B+3}{2})k_\lambda^{n_B+3}}k^{n_B}\delta({\bf k}-{\bf k}^{\prime})~,
\end{eqnarray}
Since the energy momentum tensor for magnetic field is quadratic in $B$,
one needs to calculate a convolution 
in order to obtain the spectrum of $\Delta_{B}$ and $L$. 
 Although many previous works used approximated spectra,
 Ref. \cite{2008arXiv0811.0230P} obtained the exact spectra for several
 values of $n_B$. For example, the spectrum for $n_B=-2.5$ is:
\begin{eqnarray}
&&k^3\frac{\mid \Delta_{B}(k) \mid^2_{n_B=-2.5}}{{2\pi^2}}\simeq \frac{17k}{{800\pi^6}} \Biggl[ \frac{(2\pi)^{n_B+5}B_\lambda^2}{2\Gamma(\frac{n_B+3}{2})k_\lambda^{n_B+3}(a^4\rho_\gamma)} \Biggr] ^2~,\nonumber\\
&&\mid L(k) \mid^2_{n_B=-2.5}\simeq \frac{55}{51}\mid \Delta_B(k) \mid^2_{n_B=-2.5}~~.
\label{chap3:spec}
\end{eqnarray}
This spectrum is valid for scales much larger than damping scale, i.e. $k\ll k_D$.
Since $k_D$ is sufficiently large, we use Eq. (\ref{chap3:spec}) throughout this work.

\section{Deriving Initial Condition}
In previous works
\cite{2008PhRvD..78b3510F,2008PhRvD..77l3001G,2008PhRvD..77f3003G}
initial conditions are derived neglecting the matter contributions. 
This is a good approximation for the adiabatic mode in the standard model. 
However, in the existence of the magnetic fields, the compensation
mechanism between the radiation energy perturbation and the  
magnetic energy density makes the matter contributions not to be
negligible. We can understand it as follows.
The initial conditions for purely magnetic mode derived in previous
works by neglecting matter contributions are 
\cite{2008PhRvD..78b3510F,2008PhRvD..77l3001G,2008PhRvD..77f3003G}
\begin{eqnarray}
&&\Delta_\gamma=-R_\gamma\Delta_B-\frac{1}{6}(R_\nu\Delta_B-2\pi_B)k^2\tau^2~,\\
&&\Delta_\nu=-R_\gamma\Delta_B+\frac{R_\gamma}{6R_\nu}(R_\nu\Delta_B-2\pi_B)k^2\tau^2~,\\
&&\Delta_b=-\frac{3}{4}R_\gamma\Delta_B+{\cal O}(k^2\tau^2)~,\\
&&\Delta_c=-\frac{3}{4}R_\gamma\Delta_B+{\cal O}(k^2\tau^2)~,\\
&&\z={\cal O}(k^3\tau^3).
\end{eqnarray}
In the radiation dominant epoch, if one substitutes these solutions to
Eq. (\ref{chap3:5}), one sees that contributions from radiations and magnetic
fields compensates each other to obtain
\begin{eqnarray}
\z^\p+\hub\z+\frac{1}{2k}\kappa\rho a^2 R_f(\Delta_f+3\delta P_f)=\frac{3}{2k\tau^2}(-R_mR_\gamma\Delta_B\tau+{\cal O}(\tau^3))+{\cal O}(k^3\tau^2)~,\label{chap3:old}
\end{eqnarray}
where we neglected the baryon pressure. In the early epoch, $\tau \rightarrow 0$, Eq. (\ref{chap3:old}) diverges and does not
satisfy Eq. (\ref{chap3:5}).
This inconsistency is caused by the matter contribution in r.h.s. of
Eq. (\ref{chap3:old}). 
Thus it is important to derive initial conditions including baryon and
CDM perturbations. The situation is similar to those of isocurvature
models, in which the metric perturbations at early times are determined
by the non-relativistic matter contributions \cite{2000PhRvD..62h3508B}.

In order to obtain the appropriate initial conditions with magnetic
fields for $\Phi$, $\s$ and $\z$ up to the leading order in $k\tau$, 
let us first combine Eqs. (\ref{chap3:1})-(\ref{chap3:4}). 
Although the matter contributions make equations complicated, we obtain
the following three equations in the radiation dominant epoch:
a second-order equation for $\Phi$, 
\begin{eqnarray}
&&(1+R_m\tau)\Phi^\pp+(4+3R_m\tau)\hub\Phi^\p+\frac{1}{3}\Bigl(k^2-(2\hub^\p+\hub^2)R_m\tau\Bigr)\Phi+{\cal O}(\hub)\s\nonumber\\
&&=\frac{3}{2k^2}R_f\pi_f\Bigl(-2\hub^4+2\hub^2\hub^\p+2\hub\hub^\pp+2\hub^{\p2}+(-\hub^4+2\hub^{\p2}+2\hub\hub^{\pp})R_m\tau\Bigr)\nonumber\\
&&+\frac{3}{2k^2}R_f\pi_f^\p\Bigl(2\hub^3+4\hub\hub^\p+(\hub^3+4\hub\hub^\p)R_m\tau\Bigr)+\frac{3}{2k^2}R_f\pi_f^\pp(1+R_m\tau)\hub^2\nonumber\\
&&+\frac{\hub^2}{2}(3+2R_m\tau)R_f\pi_f-\frac{\hub^2}{2}R_m\tau R_\gamma\Delta_B,\label{chap3:6}
\end{eqnarray}
a third-order equation for $\s$,
\begin{eqnarray}
&&(1+R_m\tau)\s^\ppp+(5+4R_m\tau)\hub\s^\pp+\Bigl(4\hub^\p+6\hub^2+(4\hub^2+4\hub^\p)R_m\tau+\frac{k^2}{3}\Bigr)\s^\p\nonumber\\
&&+\Bigl(2\hub^\pp+6\hub\hub^\p+(2\hub^\pp+4\hub\hub^\p)R_m\tau+\frac{k^2}{3}\hub\Bigr)\sigma\nonumber\\
&&=-\hub^2(2+R_m\tau)kR_f\pi_f+\frac{\hub^2}{2}R_m\tau kR_\gamma\Delta_B\nonumber\\
&&-\Bigl(4\hub^{\p2}+4\hub\hub^\pp+12\hub^2\hub^\p+(4\hub^{\p2}+4\hub\hub^\pp+8\hub^2\hub^\p)R_m\tau\Bigr)\frac{3}{2k}R_f\pi_f\nonumber\\
&&-\Bigl(8\hub\hub^\p+6\hub^3+(8\hub\hub^\p+4\hub^3)R_m\tau\Bigr)\frac{3}{2k}R_f\pi_f^\p\nonumber\\
&&-2(1+R_m\tau)\hub^2\frac{3}{2k}R_f\pi_f^\pp,\label{chap3:8}
\end{eqnarray}
and a first-order equation for $\z$, $\s$ and $\Phi$,
\begin{eqnarray}
\z^\p+\frac{3+2R_m\tau}{1+R_m\tau}\hub\z+\frac{2+R_m\tau}{1+R_m\tau}k\Phi-\frac{2+R_m\tau}{1+R_m\tau}\hub\s=\frac{3\hub^2}{2k}\frac{2+R_m\tau}{1+R_m\tau}R_f\pi_f-\frac{3\hub^2}{2k}\frac{R_m\tau}{1+R_m\tau}R_\gamma\Delta_B~,
\label{chap3:7}
\end{eqnarray}
where we have neglected the baryon pressure and used the adiabatic condition, $\Delta_\gamma=\Delta_\nu=\frac{4}{3}\Delta_b=\frac{4}{3}\Delta_c$. 
As we have already pointed out, $\Delta_B$ and $\pi_B$ are constant. 
The magnetic mode is a particular solution of the linearized Einstein
equations, while the standard adiabatic and isocurvature modes are
the general solutions of them.

In the purely magnetic mode, photons and neutrinos compensate the energy
perturbation of the magnetic field initially,
i.e. $\Delta_\gamma=\Delta_\nu=-R_\gamma\Delta_{B}+{\rm
(higher~order~terms)}$ \cite{2008PhRvD..78b3510F,2008PhRvD..77l3001G,2004PhRvD..70l3507G,2008PhRvD..78l3001Y},
and anisotropic stress of neutrinos compensates that of the magnetic
field, i.e. 
$\pi_\nu=-\frac{R_\gamma}{R_\nu}\pi_{B}+\pi^{(2)}k^2\tau^2$, where
$\pi^{(2)}$ denotes the coefficient of the ${\cal O}(k^2 \tau^2)$ term.
These compensation mechanism and Eqs.(\ref{chap3:6})-(\ref{chap3:7}) make it possible to derive $\Phi$, $\s$ and $\z$
up to the second order:
\begin{eqnarray}
&&\Phi=\frac{9}{2}R_\nu\pi^{(2)}-\frac{R_m}{8k}R_\gamma\Delta_Bk\tau~,
\label{chap3:13}\\
&&\s=-3R_\nu\pi^{(2)}k\tau+\frac{R_m}{24k}R_\gamma\Delta_Bk^2\tau^2~,
\label{chap3:14}\\
&&\z=-\frac{R_m}{2k}R_\gamma\Delta_B+\Bigl(-3R_\nu\pi^{(2)}+\frac{1}{8}\Bigl(\frac{R_m}{k}\Bigr)^2R_\gamma\Delta_B\Bigr)k\tau~,
\label{chap3:15}
\end{eqnarray}
where we take into account the matter contribution to the Friedmann
equation, namely, $a \simeq \sqrt{\frac{\kappa
(\rho_{\nu 0}+\rho_{\gamma
0})}{3}}\tau+\frac{\kappa(\rho_{b0}+\rho_{c0})}{12}\tau^2$ with
$\rho_{i0}$ being the energy density of species $i$ at present, 
and $\hub\simeq \tau^{-1}+\frac{1}{3}R_m$ in the radiation dominant
epoch \cite{2000PhRvD..62h3508B,2004PhRvD..70d3011L}.

In order to know initial conditions for each component, we need to solve Eqs. (\ref{chap3:b1})-(\ref{chap3:c0}).
The leading order terms of the neutrino perturbation are determined from the 
compensation mechanism.  From Eqs. (\ref{chap3:15}) and (\ref{chap3:14}), we obtain
the neutrino perturbation up to the second order:  
\begin{eqnarray}
&&\Delta_\nu=-R_\gamma\Delta_B+\frac{2}{3}R_mR_\gamma\Delta_B\tau~,\label{chap3:21}\\
&&q_\nu=-\frac{1}{3}\left(R_\gamma\Delta_B-2\frac{R_\gamma}{R_\nu}\pi_B\right)k\tau+\frac{1}{9}R_mR_\gamma\Delta_B k\tau^2~,\label{chap3:22}\\
&&\pi_\nu=-\frac{R_\gamma}{R_\nu}\pi_B-\left(\frac{2}{15}R_\gamma\Delta_B-\frac{11}{21}\frac{R_\gamma}{R_\nu}\pi_B+\frac{8}{5}R_\nu\pi^{(2)}\right)\frac{1}{2}k^2\tau^2~,\label{chap3:23}\\
&&G^{(3)}_\nu=-\frac{3}{7}\frac{R_\gamma}{R_\nu}\pi_B k\tau~.\label{chap3:24}
\end{eqnarray}
Again, the definition of $\pi^{(2)}$ is $\pi_\nu=-\frac{R_\gamma}{R_\nu}\pi_{B}+\pi^{(2)}k^2\tau^2$. Then Eq. (\ref{chap3:23}) leads to
\begin{eqnarray}
\pi^{(2)}=-\frac{1}{42}\frac{R_\gamma}{R_\nu}\frac{14R_\nu\Delta_B-55\pi_B}{4R_\nu+5}~.\label{chap3:25}
\end{eqnarray}
The equation of motion for photon-baryon fluid, Eq. (\ref{chap3:photb}), imply that
\begin{eqnarray}
&&q_\gamma=\frac{1}{3}(R_\nu\Delta_B-2\pi_B) k\tau+q_\gamma^{(2)}k\tau^2~.\label{chap3:30}
\end{eqnarray}
From Eqs.(\ref{chap3:3}) and (\ref{chap3:15}) and tight coupling approximation, $q_b\simeq 3q_\gamma/4$, we can obtain $q^{(2)}_\gamma$:
\begin{eqnarray}
&&R_fq_f
=\left(\frac{1}{9}R_mR_\nu R_\gamma\Delta_B+R_\gamma q_\gamma^{(2)}+\frac{1}{4}R_b(R_\nu\Delta_{B}-2\pi_{B})\right) k\tau^2
=\frac{1}{9}R_mR_\gamma\Delta_B k\tau^2~,\nonumber\\
&&q_\gamma^{(2)}=\frac{1}{9}R_mR_\gamma\Delta_{B}-\frac{1}{4}\frac{R_b}{R_\gamma}(R_\nu\Delta_{B}-2\pi_{B})\label{chap3:31}.
\end{eqnarray}
The Boltzmann equations and the solutions of $\z$, $q_\nu$ and
$q_\gamma$ give $\Delta_\nu$, $\Delta_\gamma$ and $\Delta_c$ up to 
the third order:
\begin{eqnarray}
&&\Delta_\nu=-R_\gamma\Delta_B+\frac{2}{3}R_mR_\gamma\Delta_B\tau+\Bigl(-2\pi^{(2)}R_\nu-\frac{1}{12}\Bigl(\frac{R_m}{k}\Bigr)^2R_\gamma\Delta_B+\frac{1}{6}(R_\gamma\Delta_B-2\frac{R_\gamma}{R_\nu}\pi_B)\Bigr)k^2\tau^2~,\label{chap3:33}\\
&&\Delta_\gamma=\frac{4}{3}\Delta_b=-R_\gamma\Delta_B+\frac{2}{3}R_mR_\gamma\Delta_B\tau+\Bigl(-2\pi^{(2)}R_\nu-\frac{1}{12}\Bigl(\frac{R_m}{k}\Bigr)^2R_\gamma\Delta_B-\frac{1}{6}(R_\nu\Delta_B-2\pi_B)\Bigr)k^2\tau^2~,\label{chap3:34}\\
&&\Delta_c=-\frac{3}{4}R_\gamma\Delta_B+\frac{1}{2}R_mR_\gamma\Delta_B\tau+\Bigl(-\frac{3}{2}\pi^{(2)}R_\nu-\frac{1}{16}\Bigl(\frac{R_m}{k}\Bigr)^2R_\gamma\Delta_B\Bigr)k^2\tau^2~.\label{chap3:34-2}
\end{eqnarray}
In the numerical calculation, we use the curvature perturbation $\eta$, $\eta=-(2\Phi+\sigma^\p/k)$,
in place of $\Phi$. Its initial condition is 
\begin{eqnarray}
\eta=-6R_\nu\pi^{(2)}+\frac{1}{6}\frac{R_m}{k}R_\gamma\Delta_{B}k\tau-\frac{1}{48}\Bigl(\frac{R_m}{k}\Bigr)^2R_\gamma\Delta_Bk^2\tau^2~.
\end{eqnarray}

Since this magnetic mode is a particular solution of the linearized
Einstein equations, the general solution can be the sum of standard
adiabatic mode and purely magnetic mode. 
In such cases, the total temperature perturbation $\Delta^{\rm tot}$ is
represented as  
\begin{eqnarray}
&&\Delta^{\rm tot}=\Delta^{\rm adi}+\Delta^{\rm B}~.
\end{eqnarray}
Here $\Delta^{\rm adi}$ is a temperature perturbation from the
adiabatic mode, which is calculated with standard adiabatic initial
conditions \cite{1995ApJ...455....7M}, and $\Delta^{\rm B}$ is from the purely magnetic mode.
Then the ensemble average is
\begin{eqnarray}
&&\langle \Delta^{\rm tot}\Delta^{\rm tot\ast}\rangle=\langle\Delta^{\rm adi}\Delta^{\rm adi\ast}\rangle +\langle\Delta^{\rm B}\Delta^{\rm B\ast}\rangle + \langle\Delta^{\rm adi}\Delta^{\rm B\ast}\rangle+ \langle\Delta^{\rm B}\Delta^{\rm adi\ast}\rangle~,
\end{eqnarray}
where the latter two terms are the correlation between the adiabatic and
the purely magnetic modes.
In what follows we study the three cases, namely, fully correlated case,
anti correlated case and uncorrelated case.

\section{Result and Discussion}
\subsection{Matter Contribution}
The new initial conditions derived in the above are totally different
from those used in the previous works. 
Since we do not neglect matter contribution, $\Delta$'s and $\eta$ have terms of ${\cal O}(R_m\tau)$.
To see the importance of $R_m\tau$ terms, we 
calculate the total energy perturbation up to the second order:  
\begin{eqnarray}
R_f\Delta_f&=&-\frac{1}{3}R_mR_\gamma\Delta_B\tau+\left(-2R_\nu\pi^{(2)}k^2+\frac{7}{12}R_m^2R_\gamma\Delta_B\right)\tau^2  \label{chap3:35-1}\\
&=&\Bigl(-\frac{1}{3}R_\gamma\Delta_B+\frac{7}{12}R_mR_\gamma\Delta_B\tau\Bigr)R_m\tau-2R_\nu\pi^{(2)}k^2\tau^2~.\label{chap3:35-2}
\end{eqnarray}
We define a wavenumber $k_{\rm mat}$ as the scale in which absolute values of two factors in ${\cal O}(\tau^2)$ term in 
Eq. (\ref{chap3:35-1}) are equal to each other,
\begin{eqnarray}
\mid 2R_\nu\pi^{(2)}k_{\rm mat}^2\mid = \Bigl| \frac{7}{12}R_m^2R_\gamma\Delta_B\Bigr|~,
\end{eqnarray}
then we obtain $k_{\rm mat}\simeq 1.2R_m$. If $k<k_{\rm
mat}$, $R_f\Delta_f$ is dominated by the matter contribution.  
Since the wave number $k_{\rm mat}$ correspond to the multipole $l_{\rm mat}\sim k_{\rm mat}\tau_0\sim 80$, where 
$\tau_0$ is the present conformal time, we cannot neglect the matter contribution in lower multipoles.
When $k>k_{\rm mat}$,
we define $\tau_{m}$ as the conformal time when the absolute values of two terms in r.h.s. of Eq.(\ref{chap3:35-2}) are equal to each other, i.e.
\begin{eqnarray}
k\tau_m\equiv  \frac{4R_mR_\gamma\Delta_{B}k}{36\mid \pi^{(2)}\mid R_\nu k^2+7R_m^2R_\gamma\Delta_{B}}\simeq \Bigl(2\frac{k}{R_m}+1.75\frac{R_m}{k}\Bigr)^{-1}~, \label{chap3:36}
\end{eqnarray}
where we assumed $1\gg k\tau>k_{\rm mat}\tau\sim R_m\tau$.
When $k\tau$ is smaller than $k\tau_m$, $R_m\tau$ term in
Eq. (\ref{chap3:35-2}), which comes from the matter contributions, 
can not be negligible and should be incorporated appropriately. 
In the radiation dominant epoch, $R_m$ is represented as
\begin{eqnarray}
R_m&\simeq&\frac{3}{4}\frac{\Omega_bh^2+\Omega_ch^2}{\sqrt[]{\mathstrut} \Omega_\gamma h^2+\Omega_\nu h^2}\frac{100{\rm ~km~sec^{-1}~Mpc^{-1}}}{\rm Mpc^{-1}}{\rm ~~(Mpc^{-1})}\nonumber\\
&\simeq&{5\times 10^{-3} {\rm Mpc}^{-1} \frac{\Omega_bh^2+\Omega_ch^2}{0.02+0.11}\Bigl(
 \frac{\Omega_\gamma h^2+\Omega_\nu
 h^2}{4.3\times10^{-5}}\Bigr)^{-1/2}}~.
\end{eqnarray}
Then we obtain $k\tau_m$ as shown in Fig. \ref{chap3:fig0}.
In small scales, $k\tau_m$ become smaller and we can neglect the matter contribution. 
However, if we start to integrate equations from the conformal time much
smaller than  
$\tau_m$, the matter contribution plays an important role in
the magnetic mode. For example, if one sets initial conditions at $k\tau
\approx 0.01$, then from Fig. \ref{chap3:fig0} we find that the modes
$k\lesssim 0.3$ suffer from the matter contributions. These modes
correspond to the angular scale $\ell \lesssim 4000$, at which the
difference should be significant as shown in Fig. \ref{chap3:fig0}.

\subsection{Numerical Calculation}
The equations and initial conditions derived in the previous sections
can be used to calculate CMB and matter power spectra numerically. We
calculated them with accordingly modified CAMB code \cite{Lewis:1999bs}. 
In all of our calculation we fixed cosmological parameters to the
best-fitting values to the WMAP-5yr data \cite{2008arXiv0803.0547K},
namely $(\omega_b, \omega_c, h, \tau_c, \Delta_R^2, n_s)=(0.0227,
0.1099,0.719,0.087,2.41\times 10^{-9},0.963)$, where $\omega_b$
and $\omega_c$ are the energy densities of baryon and CDM, respectively,
$h$ is the Hubble parameter, 
$\tau_c$ is the
optical depth, $\Delta_R^{2}$ and $n_s$ are the amplitude and the
spectral index of primordial curvature fluctuations, respectively.

\begin{figure}
\includegraphics[width=6cm,angle=270]{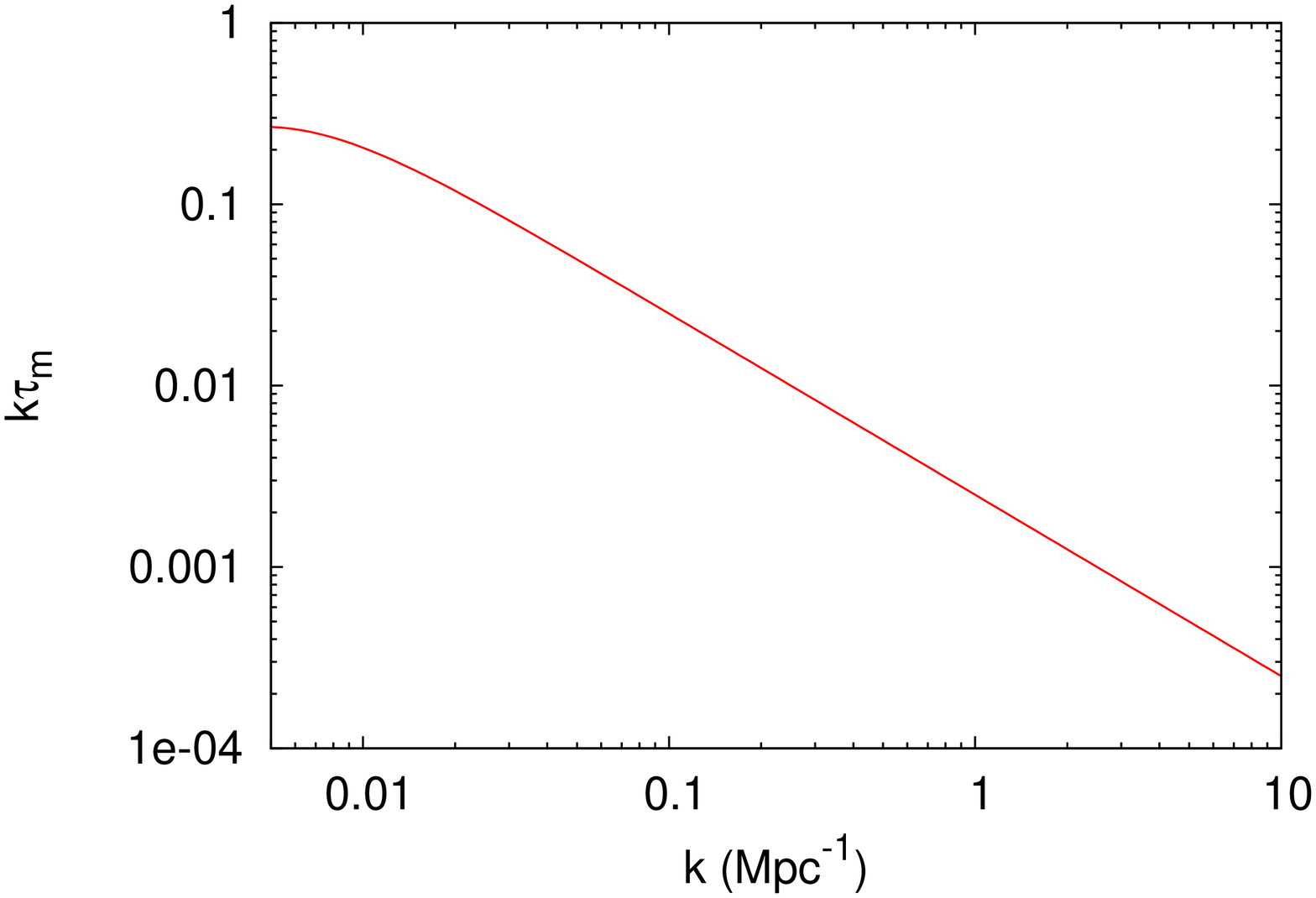}
\includegraphics[width=6cm,angle=270]{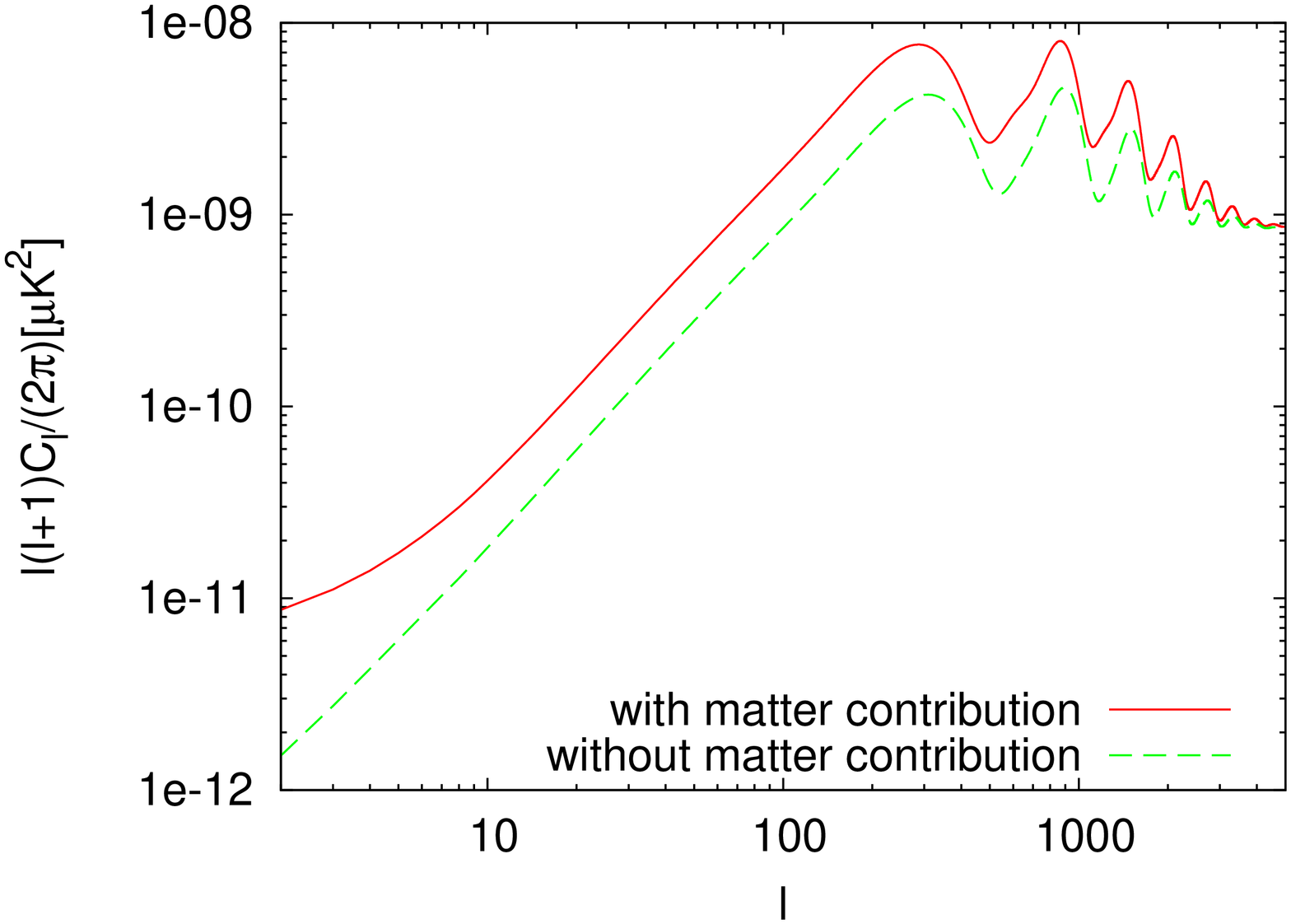}
\caption{Left: The conformal time at which the matter contribution
 becomes significant, $k\tau_m$. If $k\tau<k\tau_m$, 
 the matter contribution can not be neglected.
Right: Comparison of the CMB power spectra derived from the initial
 condition with matter contribution and from the one without. The new
 initial condition leads the larger amplitude. Magnetic field
 parameters are $B_\lambda=1{\rm nG}$ and $n_B=-2.5$.}  
\label{chap3:fig0}
\end{figure}

In Fig. \ref{chap3:fig0}, we compare the CMB spectrum of purely magnetic mode
calculated with our new initial conditions and that with previous
initial conditions, in which the matter contributions were omitted.
At higher multipoles, two spectra converge with each other
asymptotically. This is because 
the matter contributions are negligible in very small scales as shown in the previous subsection.
However, at large angular scales, we cannot neglect the matter contribution and the correct initial
condition yields larger amplitude.

The perturbed Einstein equation gives four equations and two out of four 
are independent. In order to check the consistency of our numerical
calculation, we picked up six kinds of different pairs of two equations
from the four independent equations and observed that all of the results coincide
with each other.  Note that the old initial conditions do not satisfy
the Einstein equation. We found that different pairs of 
perturbed Einstein equations yield different results if we start with the
old initial conditions.

\begin{figure}
\includegraphics[width=6cm,angle=270]{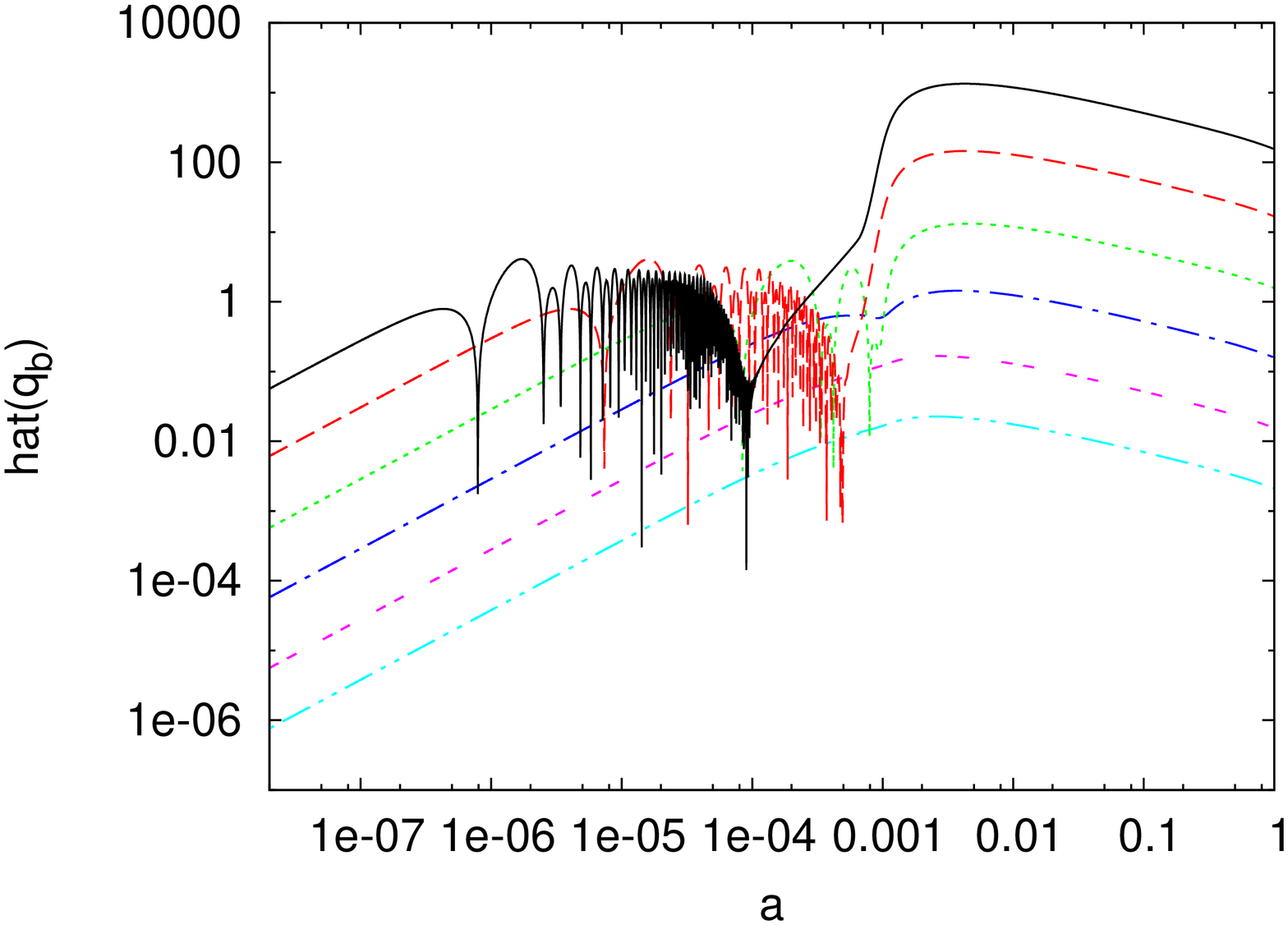}
\includegraphics[width=6cm,angle=270]{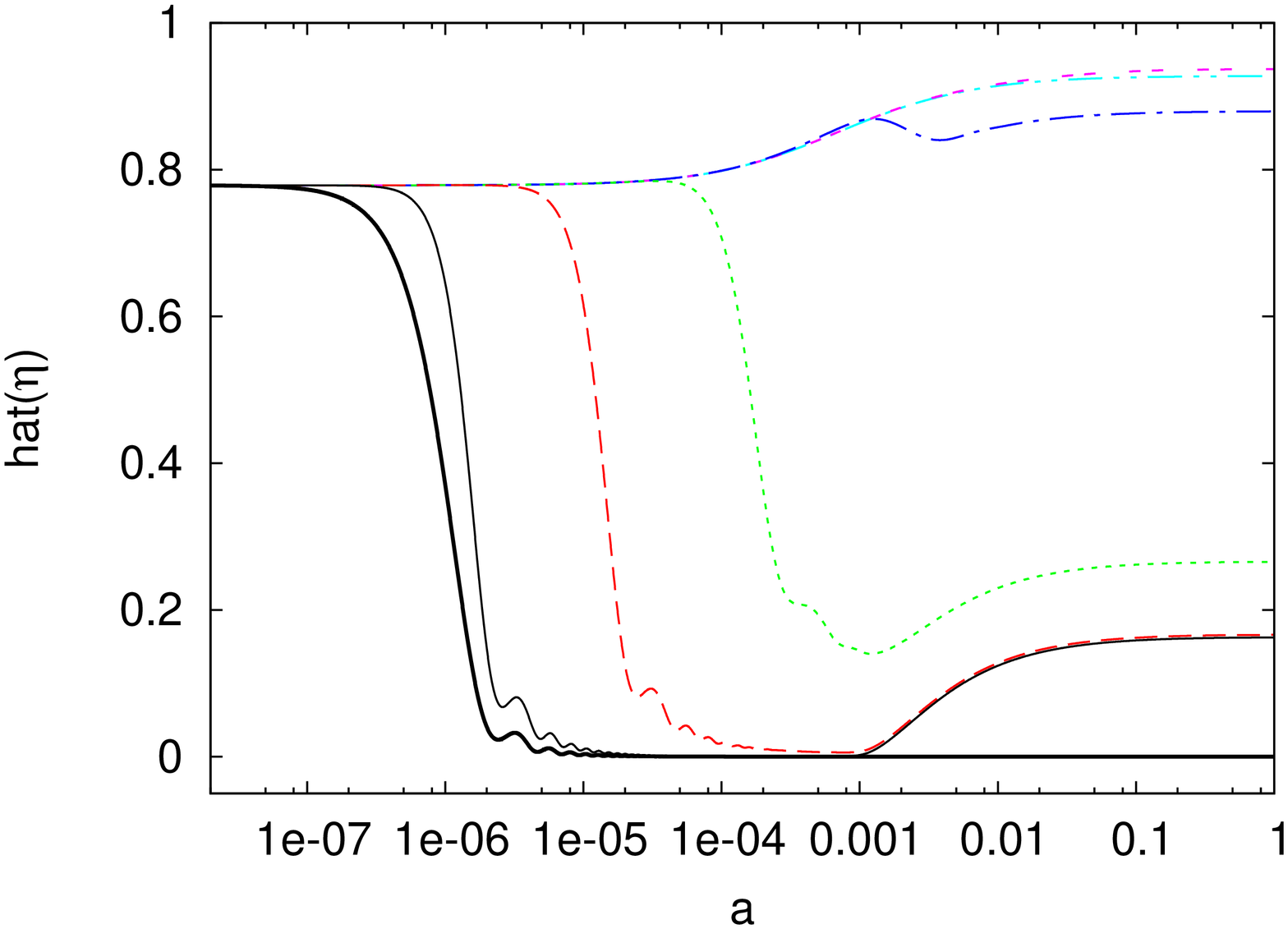}
\caption{The normalized baryon heat flux $\hat{q}_b$ (left) and
 curvature perturbation $\hat{\eta}$ (right) of the purely magnetic mode. 
The baryon velocity evolves by the Lorentz force after
 recombination. This enhances the curvature potential at 
small scales. 
Each thin line represents the purely magnetic mode with different scale: 
solid (black), dashed (red), dotted (green), dash-dotted (blue), double dotted (magenta),
 and dashed-double dotted (light blue) lines corresponds to the modes
 with $k=10$, $1$, $10^{-1}$, $10^{-2}$, $10^{-3}$, and $10^{-4}$
 Mpc$^{-1}$, respectively.
In the right panel, a thick solid line represents the standard adiabatic
 mode with $k=10{\rm Mpc^{-1}}$, which decays to zero in the radiation
 dominated era and stays constant in the matter dominated era. 
}
\label{chap3:fig1}
\end{figure}

In Fig. \ref{chap3:fig1}, we plotted the baryon heat flux $q_b$ and curvature
perturbation $\eta$ normalized by the square root of the power spectrum,
$\sqrt[]{\mathstrut \mid\Delta_B\mid^2}$,  
$\hat{q}_b\equiv q_b/\sqrt[]{\mathstrut \mid\Delta_B\mid^2}$ 
and $\hat{\eta}\equiv \eta/\sqrt[]{\mathstrut \mid\Delta_B\mid^2}$.
The characters of the growth of perturbations are different between
$k>k_{\rm rec}$ and $k<k_{\rm rec}$, where $k_{\rm rec}$ is a wavenumber 
which crosses the horizon at the recombination epoch.
In the case of $k>k_{\rm rec}$, the perturbation enters the horizon before
the recombination. After the horizon-crossing, 
the growth of the baryon velocity is suppressed and shows oscillatory
behavior by the photons pressure through Thomson scatterings.
However, after the recombination, $q_b$ grows suddenly by the Lorentz force.
This $q_b$ evolution enhances $\eta$ because the source of the curvature
perturbation is the total velocity field, i.e.
\begin{eqnarray}
\eta^\p=\frac{1}{2k}\kappa\rho a^2R_fq_f~.
\end{eqnarray}
On the other hand, at scales where the waves enter the horizon after
recombination, $k<k_{\rm rec}$,  
the baryon velocity does not undergo the suppression and there is no
sudden growth of potentials (dash-dotted (blue) line in the left panel of Fig. \ref{chap3:fig1}).   

\begin{figure}
\includegraphics[width=6cm,angle=270]{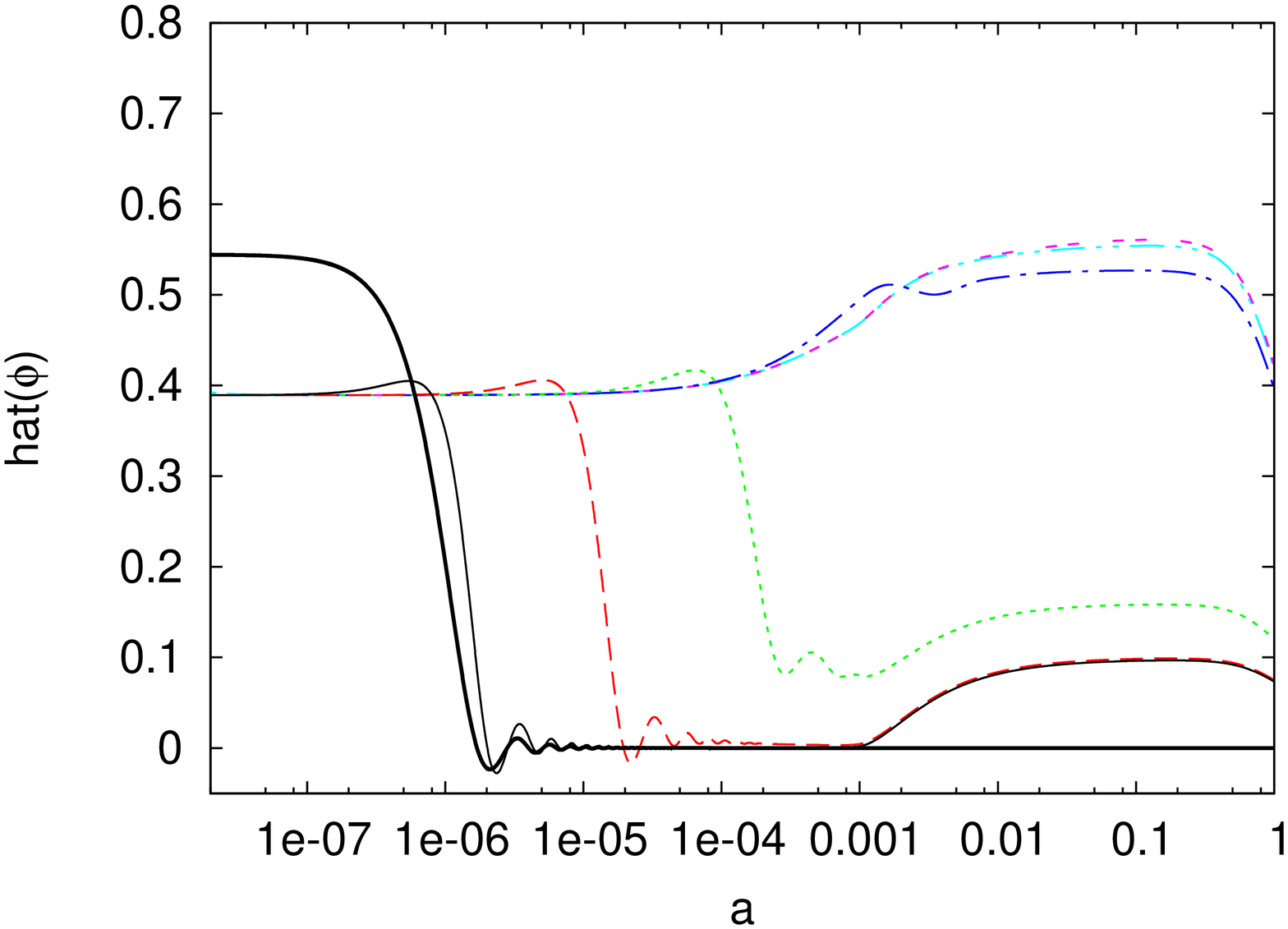}
\includegraphics[width=6cm,angle=270]{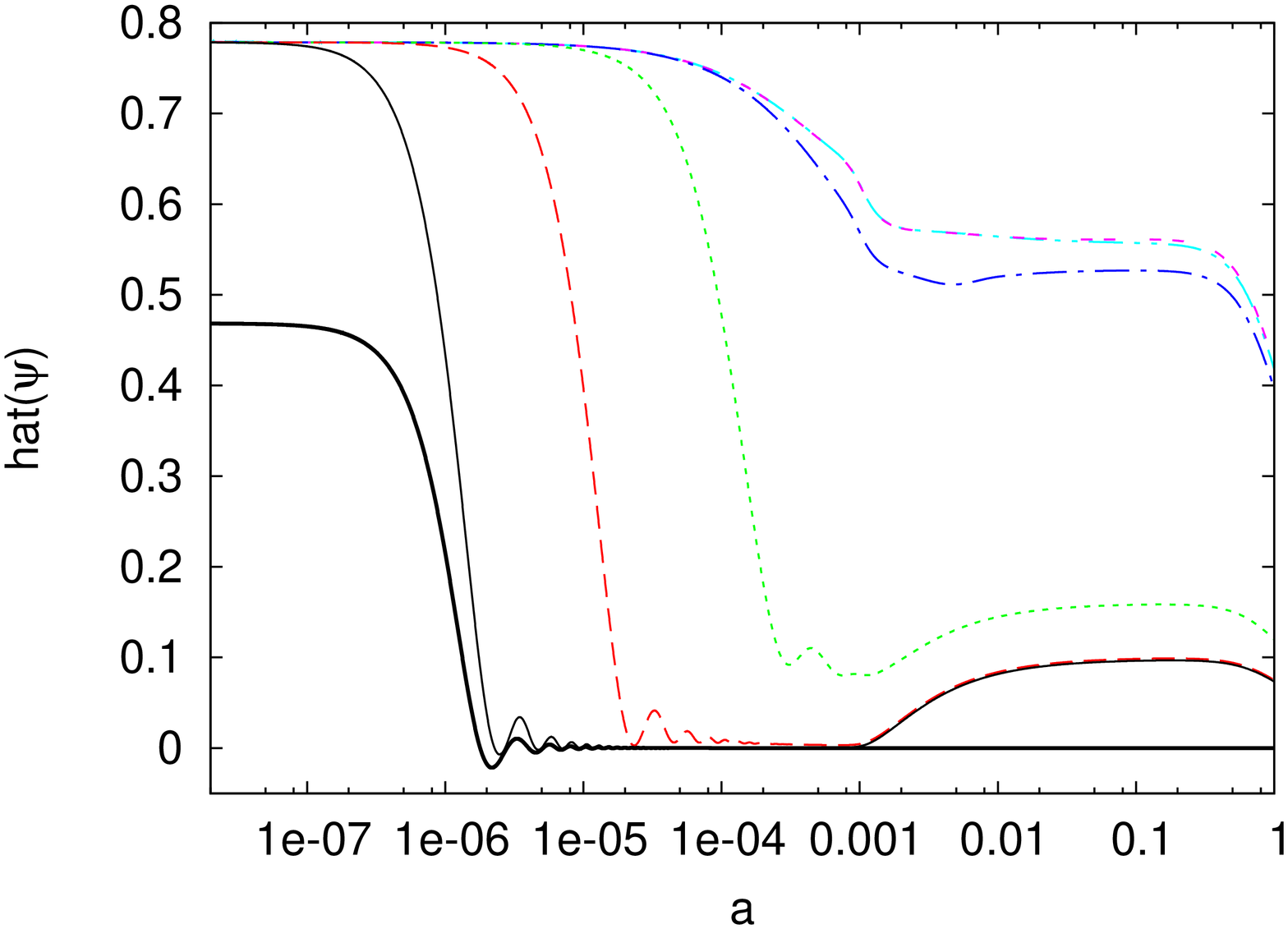}
\caption{Gravitational potentials $\hat{\phi}$ and $\hat{\psi}$ of the
 purely magnetic mode in conformal Newtonian gauge.
 Potentials grows suddenly at small scales after the recombination. 
 Each thin line represents the purely magnetic mode with different scale:
 solid (black), dashed (red), dotted (green), dash-dotted (blue),
 double dotted (magenta), and dash-double dotted (light blue) lines
 corresponds to the mode with $k=10$, $1$, $10^{-1}$, $10^{-2}$,
 $10^{-3}$, and $10^{-4}$ Mpc$^{-1}$, respectively.
For both panels, thick solid lines are the standard adiabatic modes with
 $k=10{\rm Mpc^{-1}}$.}
\label{chap3:fig2}
\end{figure}

In Fig. \ref{chap3:fig2}, two metric perturbations in conformal Newtonian
gauge are plotted.  
Again, we plotted normalized potential, $\hat{\phi}$ and $\hat{\psi}$,
with respect to the square root of the power spectrum.
Potentials $\phi$ and $\psi$ are defined as 
\begin{eqnarray}
ds^2=a^2(-(1+2\psi)d\tau^2+(1-2\phi)dx^2)~,
\end{eqnarray}
which is the same definition as in Ref. \cite{1995ApJ...455....7M}.
The relations with the variables used in this paper and their initial conditions are given as
\begin{eqnarray}
&&\phi=\eta-\frac{1}{k}\hub\sigma=-3R_\nu\pi^{(2)}+\frac{R_m}{8k}R_\gamma\Delta_Bk\tau~~,\nonumber\\
&&\psi=\frac{1}{k}(\sigma^\p+\hub\sigma)=-6R_\nu\pi^{(2)}+\frac{R_m}{8k}R_\gamma\Delta_Bk\tau~~.
\end{eqnarray}
In the absence of the magnetic field, two potentials decay to zero,
once the perturbation enter the horizon in radiation dominant epoch
(thick lines in Figs. \ref{chap3:fig1} and \ref{chap3:fig2}), and are constant in
time in matter dominated epoch.
However, in the purely magnetic mode, potentials grow after the
recombination.

\begin{figure}
\includegraphics[width=6cm,angle=270]{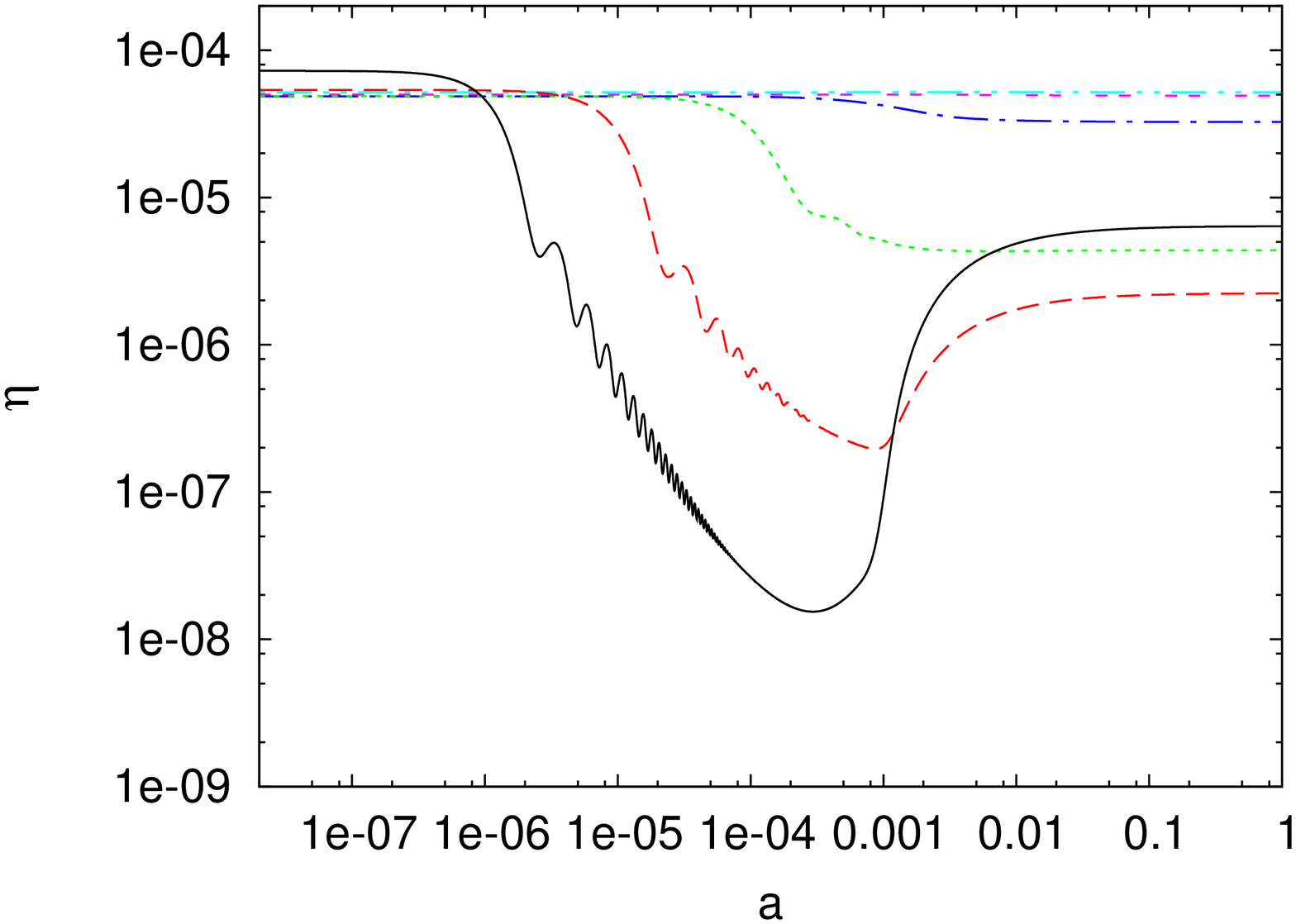}
\includegraphics[width=6cm,angle=270]{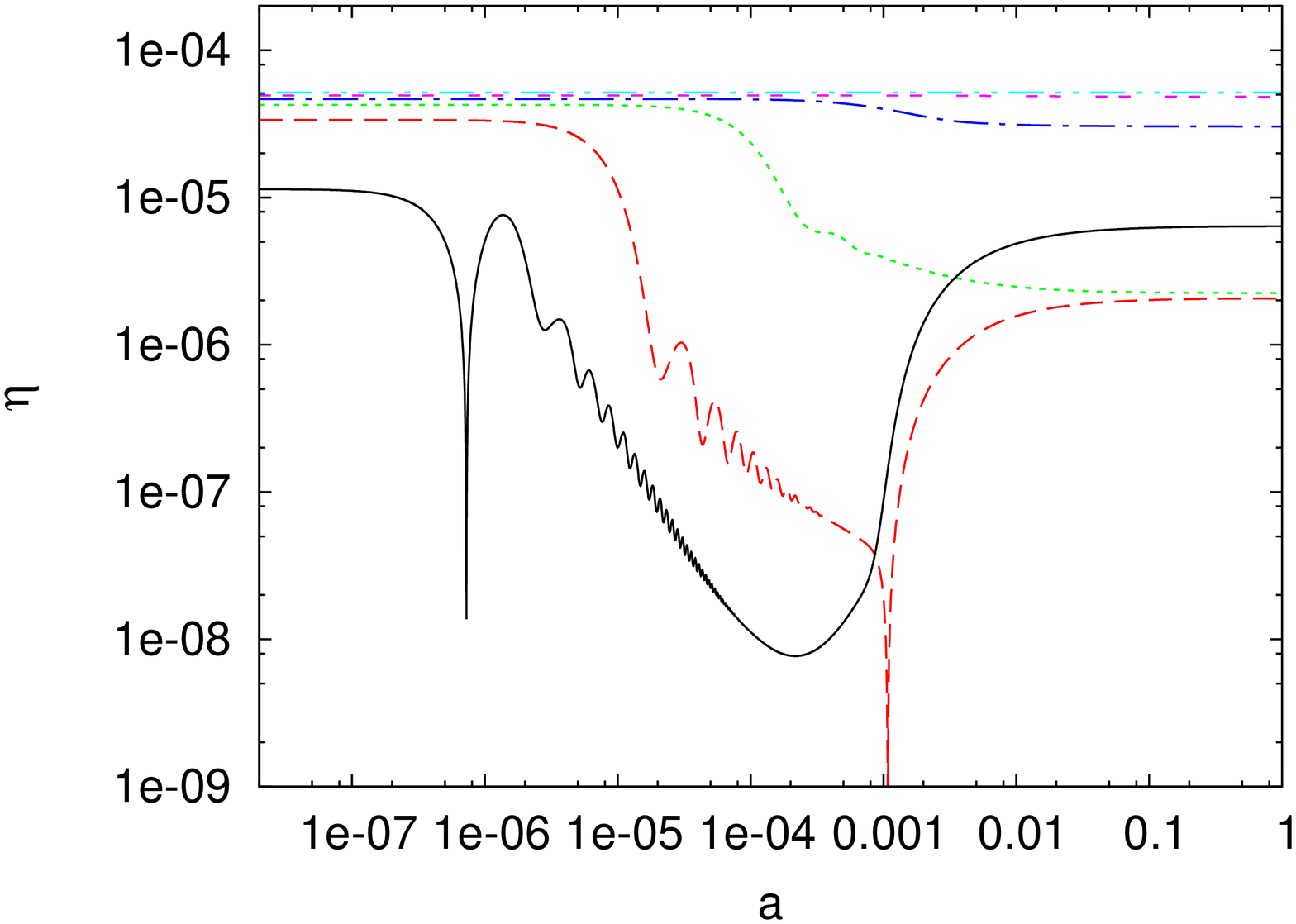}
\caption{The growths of $\eta$ in fully (left) and anti (right) 
correlated cases. 
The magnetic field enhances the curvature perturbation 
in the fully correlated case. On the other hand, anti correlated
 magnetic field diminishes $\eta$. 
Each thin line represents different scale:
solid (black), dashed (red), dotted (green), dash-dotted (blue), double
 dotted (magenta), and dash-double dotted (light blue) lines corresponds
 to the mode with $k=10$, $1$, $10^{-1}$, $10^{-2}$, $10^{-3}$, and
 $10^{-4}$ Mpc$^{-1}$, respectively.
 Parameters for magnetic fields are 
$B_\lambda=300{\rm nG}$ and $n_B=-2.5$.}
\label{chap3:fig3}
\end{figure}

\begin{figure}
\includegraphics[width=6cm,angle=270]{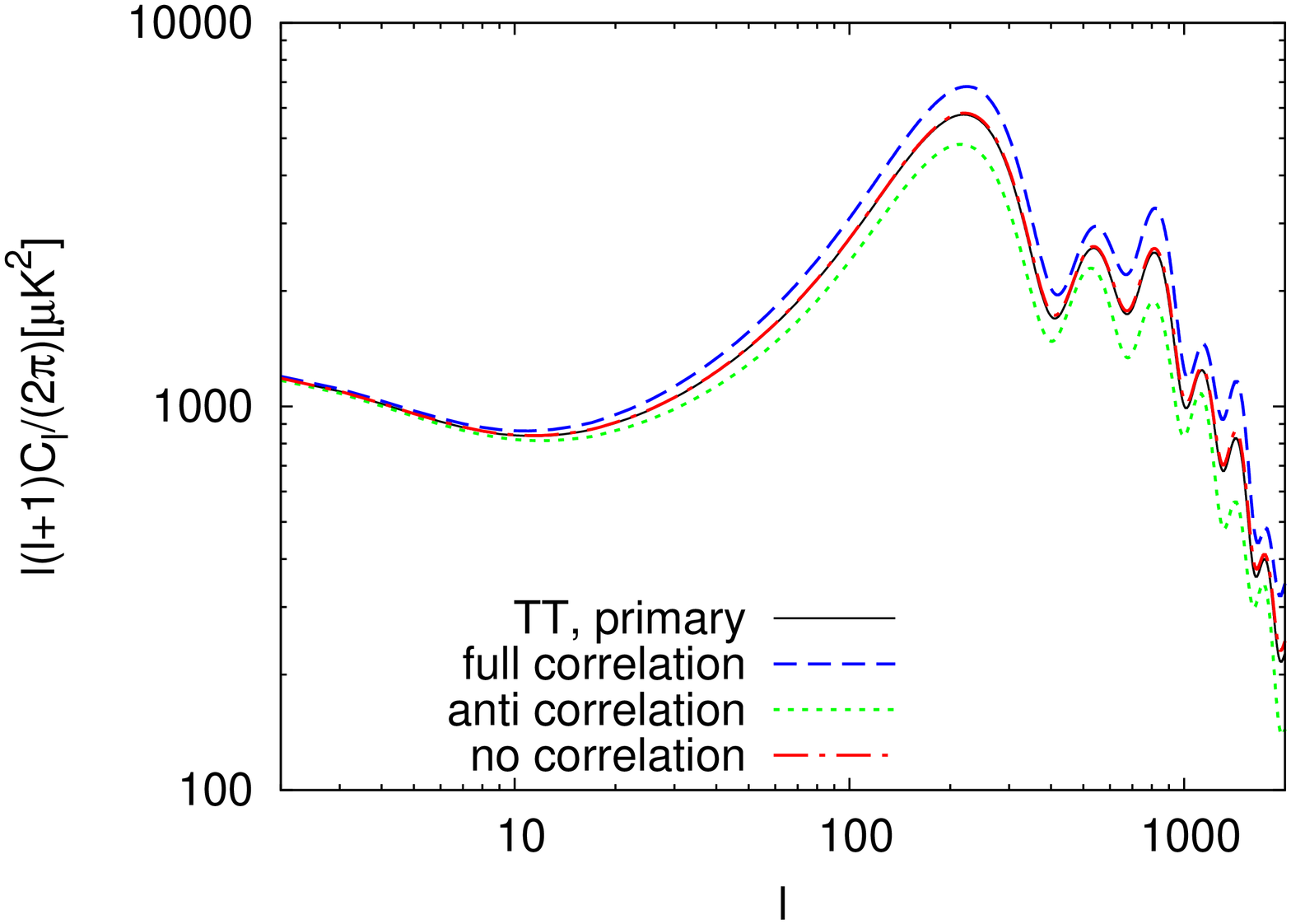}
\includegraphics[width=6cm,angle=270]{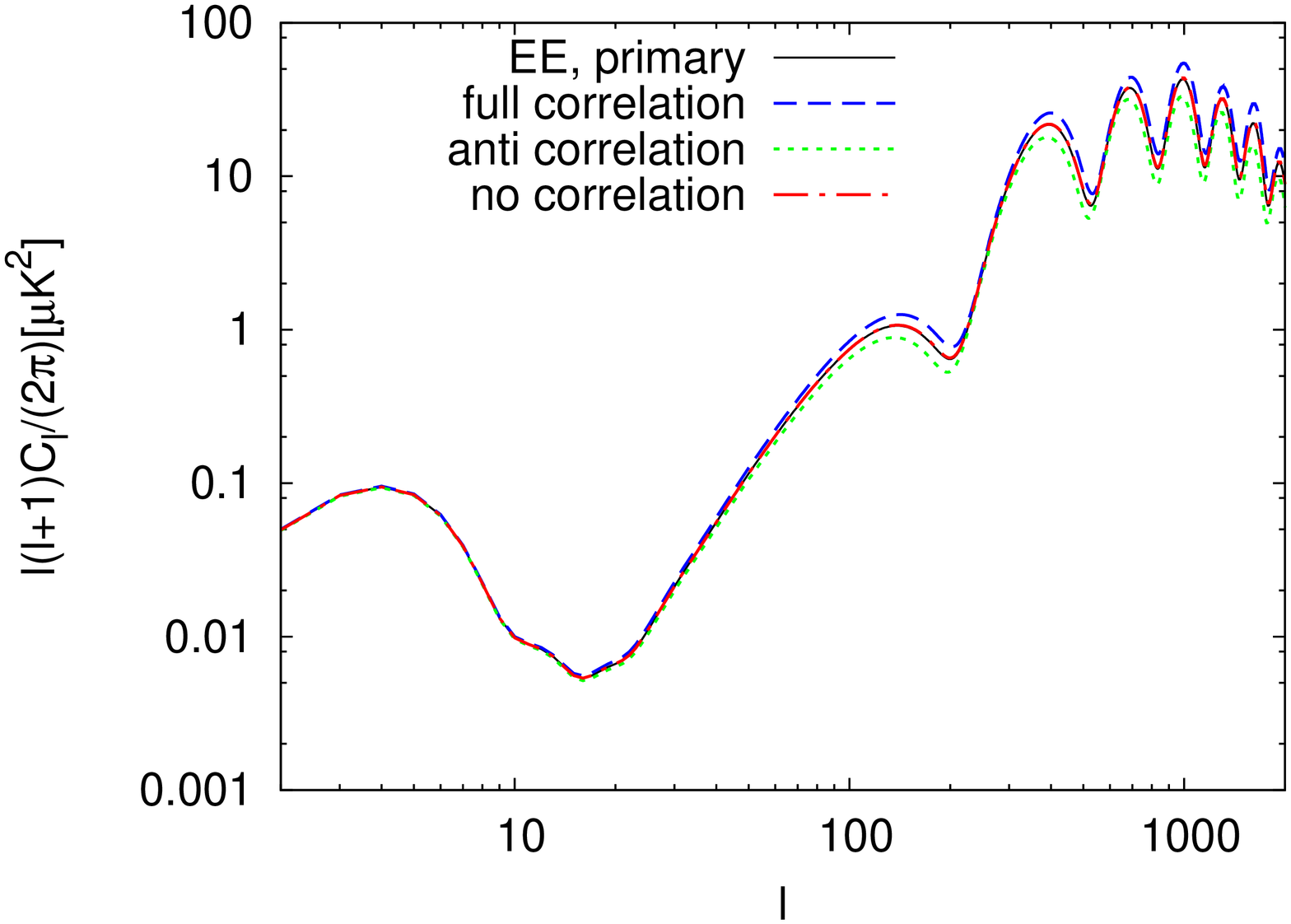}
\includegraphics[width=6cm,angle=270]{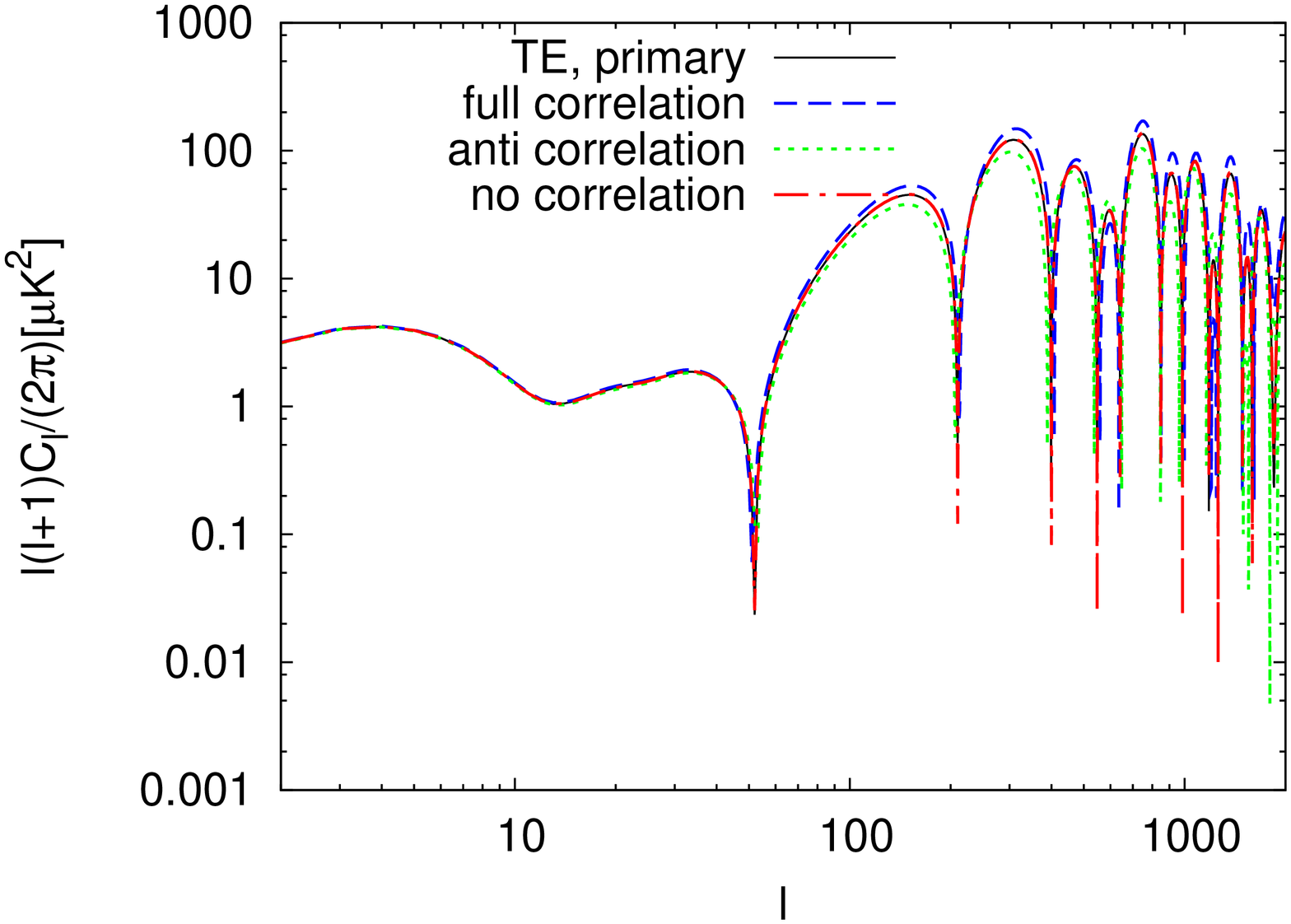}
\includegraphics[width=6cm,angle=270]{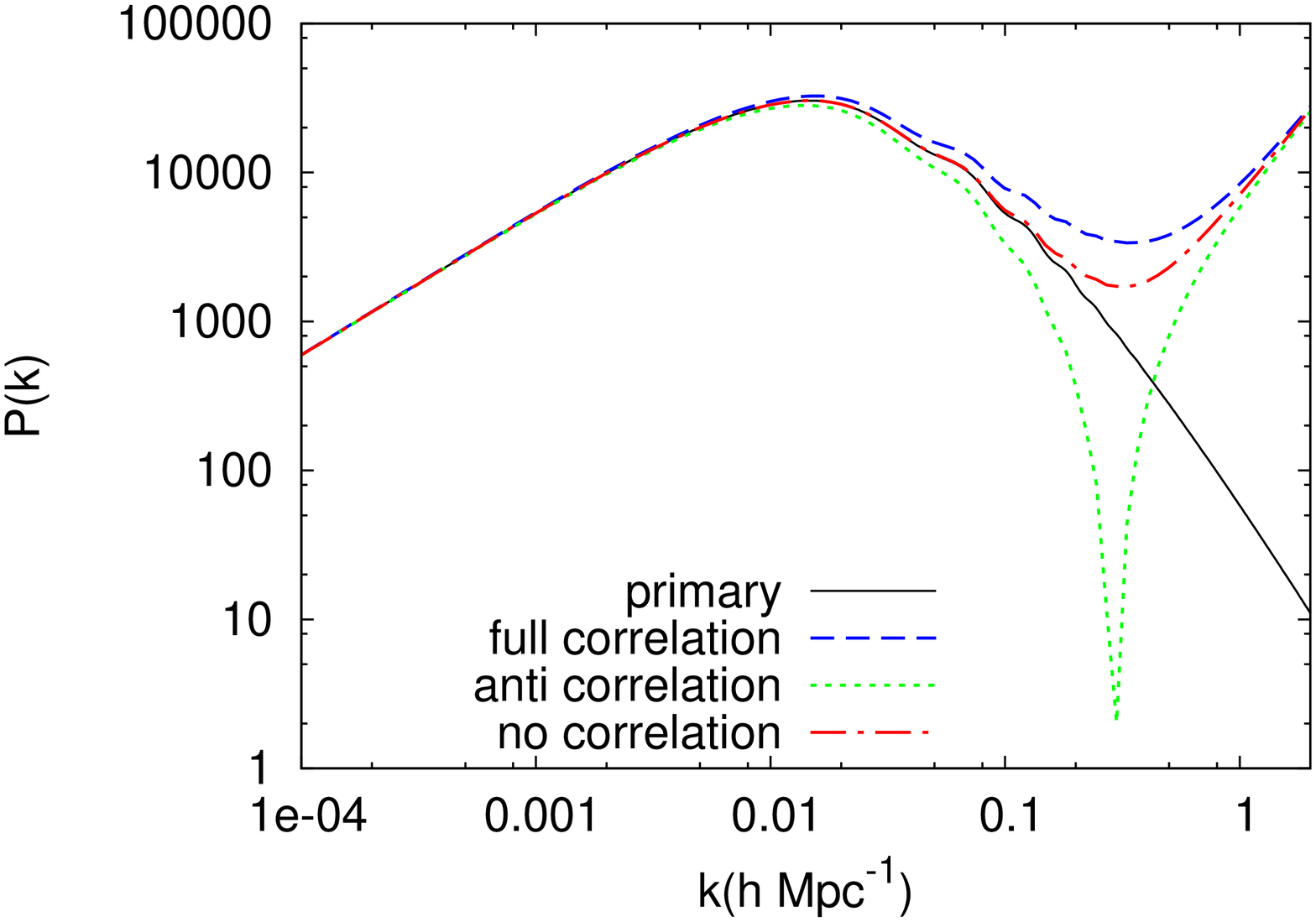}
\caption{The CMB spectra of TT, EE and TE mode and the matter power
 spectrum. If the magnetic fields of $300$ nG are uncorrelated with the
 primary adiabatic modes the effects are not significant.
 However, if the adiabatic mode is fully correlated with the
 magnetic field, the gravitational potential grows and enhances the
 spectra, which can be seen in the figures. On the other hand, the
 amplitude decreases in the anti correlated case. 
Parameters for magnetic fields are 
$B_\lambda={300}{\rm nG}$ and $n_B=-2.5$.}
\label{chap3:fig4}
\end{figure}

Next we study the growth of curvature perturbations in the adiabatic
mode correlated with the magnetic mode. 
In Fig. \ref{chap3:fig3}, the growths of $\eta$ in full and anti correlated
cases are plotted. 
If the adiabatic mode is fully correlated with the magnetic mode, $\eta$
grows after the recombination for $k>k_{\rm rec}$, while $\eta$ decays in
the anti correlated case.  
This growth of the curvature perturbation is directly seen in the 
CMB and matter power spectra as shown in Fig. \ref{chap3:fig4}. 
For the CMB spectrum (three panels in Fig. \ref{chap3:fig4}), primary standard
adiabatic mode and uncorrelated mode have a similar feature in
shape and almost degenerated. Therefore, if the adiabatic mode is fully 
correlated with the magnetic mode, the gravitational potential becomes
deeper and the amplitudes of spectra are increased.
On the other hand, the anti correlated magnetic mode decays the
potential and amplitudes of spectra become smaller. 

For the matter power spectrum, on the other hand, the difference shows
up at small scales because the Lorentz force from magnetic fields newly
induces the density perturbations dominantly at small scales after
cosmological recombination.  For both correlation cases with {$300$} nG
magnetic field, the linear power at $k\gtrsim 1$Mpc$^{-1}$ is dominated
by the perturbations induced by magnetic fields.

\section{conclusion}
In this work, we derived the initial conditions for density perturbations in the
existence of the primordial magnetic field. Because the compensation
mechanism between radiation 
perturbation and the magnetic field makes it impossible to neglect
matter contributions even in the early universe, we derived initial
conditions including matter contributions to the density field. The
initial condition derived in this paper fully satisfies the linearized
Einstein equations, and thus  
enables us to solve the system numerically in a stable and consistent manner.
Then CMB and matter
power spectra are calculated, and the evolutions of perturbations are
presented in detail. 
We found that it gives the larger amplitude of the CMB angular power
spectrum by at most a order of magnitude at large scales $l\lesssim 4000$,
compared with the initial condition in the literature.   

In the purely magnetic mode, the potentials grow suddenly after
recombination for $k>k_{\rm rec}$. This effect enhance the amplitude of the
matter power spectrum if the adiabatic and magnetic mode are fully
correlated, and decreases the amplitude in the anti correlated case at
intermediate scales. For much smaller scales $k\gtrsim 1$ Mpc$^{-1}$,
the spectrum is dominated by the magnetic mode.

%--------------------------------------------------------------------------
\acknowledgments
The authors thank T. Kajino, G. J. Mathews, and D. Yamazaki for useful
discussions. 
This work is in part supported by Grant-in-Aid for the Global Center
of Excellence program at Nagoya University `` Quest for Fundamental
Principles in the Universe:  from Particles to the Solar
 System and the Cosmos'' from the Ministry of Education, Culture,
 Sports, Science and Technology (MEXT) of Japan.

\bibliography{paper6}

%\begin{thebibliography}{99}
%\end{thebibliography}

\end{document}